\documentclass[10pt,twocolumn]{article}
\usepackage[top=2cm, bottom=2cm, left=2cm, right=2cm]{geometry}
\usepackage{times}  %
\usepackage[sort&compress,numbers]{natbib}  %
\usepackage[hyphens]{url}
\usepackage{graphicx}  %
\usepackage{flushend}
\usepackage[hyphens]{url}  %
\usepackage{graphicx} %
\usepackage{tikzsymbols}

\newcommand{\descr}[1]{\smallskip\noindent\textbf{#1}}

\usepackage{sectsty}
\sectionfont{\bfseries\Large\raggedright}

\newcommand{\pol}{/pol/\xspace}
\newcommand{\donald}{r/The\_Donald\xspace}

\newcommand{\system}{\textsc{TubeRaider}\xspace}
\newcommand{\pval}{{0.0002}\xspace}
\newcommand{\reduce}{\vspace{-0.2cm}}

\usepackage{url}                                %
\usepackage{array,multirow,graphicx,adjustbox}  %
\usepackage{booktabs}                           %
\usepackage[utf8]{inputenc}
\usepackage[ruled,algosection,noend,linesnumbered]{algorithm2e}
\usepackage{float}
\usepackage{paralist}
\usepackage{amsmath}
\usepackage{amsfonts}
\usepackage{xcolor}
\usepackage{csquotes} 
\usepackage{tabularx}
\usepackage{makecell}
\usepackage{pbox}
\usepackage[hang,flushmargin]{footmisc}
\usepackage{flushend}
\usepackage{xspace}
\usepackage{multicol, lipsum}
\usepackage[T1]{fontenc}

\usepackage{xcolor}
\usepackage{booktabs}
\usepackage{graphicx}
\usepackage{paralist}
\usepackage[small,bf]{caption}
\usepackage{subfigure}

\let\oldbibliography\thebibliography
\renewcommand{\thebibliography}[1]{%
  \oldbibliography{#1}%
  \setlength{\itemsep}{2pt}%
}

\usepackage[hang,flushmargin]{footmisc}

\usepackage[compact]{titlesec}
\titlespacing*{\section}{0pt}{*3}{3pt}
\titlespacing*{\subsection}{0pt}{*2}{2pt}
\usepackage{xspace}

\usepackage[skipabove=5pt,skipbelow=5pt,roundcorner=10pt,framemethod=TikZ]{mdframed}

\makeatletter
\def\url@leostyle{%
  \@ifundefined{selectfont}{\def\UrlFont{}}%
  {\def\UrlFont{}}%
}
\makeatother
\usepackage[hyphenbreaks]{breakurl}

\usepackage[bookmarks=true, bookmarksnumbered=true, colorlinks=true, linkcolor=linkcol, citecolor=citecol, urlcolor=urlcol, hypertexnames=true]{hyperref}

\definecolor{darkgreen}{RGB}{0, 100, 0}
\definecolor{linkcol}{rgb}{0.3,0,0}
\definecolor{citecol}{rgb}{0.3,0,0}
\definecolor{urlcol}{rgb}{0.3,0,0}

\makeatletter
\def\url@leostyle{%
  \@ifundefined{selectfont}{\def\UrlFont{\small}}%
  {\def\UrlFont{}}%
}
\makeatother
\begin{document}

\sloppy
	
	\title{\bf \system: Attributing Coordinated Hate Attacks on YouTube Videos to their Source Communities\thanks{Accepted for publication at the 18th International AAAI Conference on Web and Social Media (ICWSM 2024). Please cite accordingly.}}
	
	\author {
	Mohammad Hammas Saeed\textsuperscript{\rm 1},
	Kostantinos Papadamou\textsuperscript{\rm 2},
	Jeremy Blackburn\textsuperscript{\rm 3},\\
	Emiliano De Cristofaro\textsuperscript{\rm 2,4}, and
	Gianluca Stringhini\textsuperscript{\rm 1} \\[1ex]
	\textsuperscript{\rm 1}Boston University,
	\textsuperscript{\rm 2}University College London,
	\textsuperscript{\rm 3}Binghamton University,\\
	\textsuperscript{\rm 4}University of California, Riverside
}

\date{}
	
	\maketitle
	
	\definecolor{vlightgray}{gray}{0.925}
	
	\begin{abstract}
		Alas, coordinated hate attacks, or {\em raids}, are becoming increasingly common online.
		In a nutshell, these are perpetrated by a group of aggressors who organize and coordinate operations on a platform (e.g., 4chan) to target victims on another community (e.g., YouTube).
		In this paper, we focus on attributing raids to their source community, paving the way for moderation approaches that take the context (and potentially the motivation) of an attack into consideration.
		
		We present \system, an attribution system achieving over 75\% accuracy in detecting and attributing coordinated hate attacks on YouTube videos.
		We instantiate it using links to YouTube videos shared on 4chan's \pol board, \donald, and 16 Incels-related subreddits.
		We use a {\em peak detector} to identify a rise in the comment activity of a YouTube video, which signals that an attack may be occurring.
		We then train a machine learning classifier based on the community language (i.e., TF-IDF scores of relevant keywords) to perform the attribution.
		We test \system in the wild and present a few case studies of actual aggression attacks identified by it to showcase its effectiveness.
	\end{abstract}

	\section{Introduction}
	Coordinated hate attacks are a nefarious online phenomenon whereby bad actors organize on a social platform to orchestrate attacks disrupting other users or communities~\cite{despoina2017websci, kumar2018community, mai2018hate}.
	A high-profile example is the {\em GamerGate} campaign, when 4chan users coordinated and recruited others to raid the \#GamerGate hashtag on Twitter, targeting women and genderqueer people in the gaming industry with relentless hate speech, death threats, doxxing, and swatting attempts.
	
	Anecdotal evidence suggests that some online communities, e.g., KiwiFarms~\cite{kiwifarms2021}, are almost exclusively dedicated to orchestrating these attacks; others, e.g,  4chan~\cite{hine2017kek}, are notorious for that.
	Since this type of abuse is generated by humans rather than automated programs, techniques to detect content produced by bots~\cite{zhao2009botgraph,stringhini2015evilcohort,nilizadeh2017poised, stringhini2010detecting} are not effective. %
	
	In this paper, we focus on coordinated hate attacks on YouTube videos.
	YouTube is one of the most popular video-sharing platforms %
	and, alas, it is increasingly targeted by cyber-aggression campaigns~\cite{grigg2010cyber}.
	In fact, prior work shed light on fringe communities routinely targeting YouTube videos with raids~\cite{hine2017kek}. %
	
	\descr{Motivation.} Past research on mitigating cyber-aggression has mostly neglected the {\em attribution} of attacks, rather focusing on identifying and removing hateful messages~\cite{chen2019abuse, davidson2017AutomatedHateSpeech, djuric2015hate}, or detecting hateful users~\cite{mai2018hate, horta2018char}.
	This prompts the need to analyze coordinated aggression attacks from the lens of the online community where the attackers coordinate and the target platform where the attack occurs.
	
	Furthermore, knowing which community is responsible for a coordinated attack could assist platforms in designing more targeted mitigation techniques, e.g.,
	by factoring in context and motivation into moderation decisions. %

	\descr{Technical Roadmap.} This paper presents \system, a system that attributes coordinated hate attacks on YouTube to the online community that organized them.
	We instantiate it using data from three toxic communities: 1) 4chan's politically incorrect (\pol) board, 2) the \donald subreddit, and 3) a collection of 16 Incels subreddits identified by~\citet{papa2021over}. 
	We collect all links to YouTube videos posted on these communities and the comments on those videos on YouTube.
	
	Our analysis shows %
	that videos exhibit a peak in comment activity once they are shared on a platform, possibly indicating that a raid is taking place.
	Thus, we make \system model the comment activity and detect peaks in the comments of the video when the link is posted on a community.
	We identify important keywords for each community using TF-IDF %
	and use those as features to train a classifier.
	This approach allows us to perform a more comprehensive attribution of aggression attacks; rather than focusing on what is being said in the comments to a YouTube video, which is context-dependent and changes over time, we rely on the typical language used by hateful online communities.

	\system achieves accuracy above 75\% in attributing a coordinated attack on a given video.
	We also run \system in the wild and identify 700 videos that were likely targeted by coordinated attacks.
	We then compare the commenting activity on identified videos with those that are not attributed by \system and regular YouTube videos across a variety of inflammatory markers to demonstrate the increased toxicity, abuse, and targeted hate in raided videos.
	We also report case studies of actual attacks that exhibit hate speech and target different individuals.  %
	
	\descr{Implications.} Overall, our work paves the way for more effective content moderation.
	\system enables approaches that take into account the context and the motivation of online attacks by attributing them to the community that orchestrated them.
	This can help identify terms of service violation (e.g., attacking people based on their race or gender) and make online communities safer. 

	Due to the topic we focus on, this paper contains examples of misogynistic, hateful, and toxic content; reader discretion is advised.
	We discuss broader implication of our work, in more detail, in Section~\ref{sec:broad}.

	\section{Related Work}

	\descr{Online Aggression.} Research on online aggression mainly focuses on analyzing the toxic content posted on various platforms~\cite{chen2019abuse, djuric2015hate, davidson2017AutomatedHateSpeech}.
	\citet{salminem2019hate} detect hateful comments in the context of online news media, while \citet{alex2018hate} analyze hateful speech on Twitter and Reddit in relation to extremist violence.
	\citet{zannettou2018origins} study the spread of hateful memes on the Web with a focus on 4chan's \pol board and \donald subreddit.
	\citet{cheng2021anti} predict accounts that will engage in antisocial behavior for popular websites and detect antisocial behavior in comments, while
	\citet{chelmis2019cyber} predict if a hateful comment on an Instagram post will be followed by further hateful comments.
	\citet{jaki2019online} focus on the Incels.me forum and propose a deep learning classifier that analyzes the users' language and detects instances of misogyny, homophobia, and racism.
	Another line of work focuses on characterizing users who violate terms of service, e.g., aiming to distinguish users that post hateful content on Twitter from others~\cite{mai2018hate, horta2018char}.
	
	\descr{YouTube.} Specific to YouTube is the work by~\citet{kwon2017trump}, who analyze swearing comments against Donald Trump on YouTube.
	\citet{moor2010flaming} study ``flaming'' (i.e., the use of hostile/offensive language) on YouTube videos.
	\citet{agarwal2014crawler} detect YouTube videos {\em promoting} hatred using user and video features, while \citet{giann2010violence} detect violence using a variety of features like audio, video, and text.
	\citet{papa2021over} study the Incels community on YouTube and how to detect Incels-related videos based on a dictionary of Incel-related terms, while \citet{papadamou2022just} assess the effects of watch history on YouTube's pseudoscientific video recommendations.

	\citet{sureka2010mining} find hateful videos on YouTube via social network analysis, while \citet{weaver2012viol} show that violence in YouTube videos has more realistic consequences and a more negative nature than television violence.
	Ottoni et al.~\cite{ottoni2018right} observe that right-wing YouTube channels feature hateful and discriminatory content, \citet{wotanis2014gender} that female YouTubers receive more negative feedback in terms of sexist and hateful comments as compared to male YouTubers, and~\citet{tucker2013youtube} that a quarter of the most-viewed YouTube videos include misogynistic language, violence, or both, while the primary actors are male.

	\citet{alshamrani2020investigating} investigate the correlation of toxic behaviors like identity hate and obscenity in users' interactions with popular videos.
	\citet{tahir2019youtube} detect inappropriate videos targeting children on YouTube.
	
	Other research efforts focus on detecting cyberbullying on YouTube; e.g.,~\citet{marathe2015approach} develop a semi-automated system to identify cyber-bullying in videos, while~\citet{dadvar2014expert} use machine learning to identify cyber-bullies on YouTube.

	\descr{Coordinated Aggression.} As part of an exploratory study of 4chan, \citet{hine2017kek} shed light on raids coordinated on \pol targeting communities on Twitter and YouTube.
	This work motivates ours by showing how a video's surge in comment activity follows it being shared on another platform.

	Some work also focuses on detecting accounts involved in coordinated campaigns.
		\citet{pacheco2021coord} use an unsupervised network-based method to discover groups of accounts participating in coordinated influence campaigns, while~\citet{sharma2021coord} model account activity and hidden group behaviors to separate coordinated accounts from normal social media users.
		\citet{hernandez2018fraud} attributes fraudulent user accounts in online peer-opinion systems to a set of known fraudsters; their system takes in input a seed set of known fraudster profiles and iteratively attributes more users controlled by the same fraudster using graph deep learning.
		Conversely, our system uses TF-IDF-based approach to model the language of online communities and attributes aggression attacks on YouTube videos to a source community without focusing on whether the accounts posting comments are controlled by the same actor.
	
	\citet{mariconti2018you} use ensemble learning to predict whether a YouTube video will be raided, %
	while~\citet{kumar2018community} investigate \emph{brigading} on Reddit, whereby sub-communities form alliances and perpetrate hate crimes against competitor communities.
	By contrast, we develop a generalizable model that can assign coordinated aggressive attacks to the community that planned and executed them.

	\section{Dataset}
	We now provide an overview of the communities we study as well as the data we collect.
	Overall, we gather 1,143,988 youtube videos linked from 4chan and Reddit.

	\subsection{Background}
	
	\descr{\pol.}
	4chan is an imageboard created in 2003. %
	As of January 2023, it features 76 boards covering topics ranging from video games to Japanese culture, politics, and adult content.
	Users create new threads by posting an image to a board along with a message; other users can reply by posting messages and/or images.
	We focus on the Politically Incorrect board (\pol), which is known for the high volume of offensive content and very loose moderation.
	We do so as prior work~\cite{hine2017kek} has uncovered attempts of coordinated aggression attacks targeting YouTube users.
	
	\descr{\donald} was a subreddit created in 2015 in support of Donald Trump's 2016 U.S. Presidential Election campaign.
	It was broadly linked to the alt-right movement, and rife with racist and sexist content~\cite{donaldhatesuspend}.
	In 2019, it was quarantined and restricted before being banned.\footnote{Quarantined subreddits do not generate revenue and, among other things, require users to explicitly opt-in to viewing the content; subreddits placed in restricted mode prevent most users from creating new posts.}
	We choose this subreddit as previous work~\cite{flores2018mobilizing} analyzing the behavioral patterns of its active participants revealed frequent calls to action that ultimately lead to disruptive behavior.

	\descr{Incels Subreddits.}
	Incels (an abbreviation of Involuntary Celibates) are an online subculture of people who identify as unable to get a romantic or sexual partner despite desiring one.
	They are part of a larger collection of groups loosely organized around ``men's rights,'' known as the Manosphere~\cite{debbie2019masc}, which are often associated with promoting masculism~\cite{melissa2012masc}.
	Incels mainly focus on sexual deprivation, which they blame on their unattractive appearance.
	They believe that women are attracted to men with specific facial attributes and racial backgrounds, a lack of which leads to celibacy.
	
	We choose the Incels community as prior work studying the Manosphere~\cite{Ribeiro2020FromPA} shows it is highly engaged and produces a high volume of hateful speech.
	We start from the 19 Incels subreddits identified by~\citet{papa2021over}, but ultimately work with 16 as the remaining three do not have any YouTube links.
	
	\begin{table}[t]
		\begin{center}
			\setlength{\tabcolsep}{2pt}
			\small
			\resizebox{\columnwidth}{!}{%
				\begin{tabular}{llrrr}
					\toprule
					& & & \textbf{\#YouTube} & \textbf{\#YouTube}\\ 
					\textbf{Source}  & \textbf{Dates} & \textbf{\#Posts} &  \textbf{Links} &  {\bf Comments} \\\midrule
					\pol                & 2016--2019 & 134.5M  &    850,523  & 1B  \\ 
					\donald             & 2015--2019 & 4.8M &   278,849 & 63.7M   \\ 
					r/Braincels          & 2017--2019 & 216,806 & 7,902 & 16.4M \\
					r/ForeverAlone       & 2010--2019 & 134,723 & 4,427 & 5.3M \\
					r/Incels             & 2014--2017 & 54,218 & 2,020 & 2.9M \\
					r/IncelTears         & 2017--2019 & 61,765 & 923 & 583,057  \\
					r/IncelsWithoutHate  & 2017--2019 & 16,217 & 491 & 402,075  \\
					r/ForeverUnwanted    & 2016--2019 & 1,898 & 58  & 19,666  \\
					r/BlackPillScience   & 2018--2019 & 1,129 & 44 & 6,187   \\
					r/gymcels            & 2018--2019 & 226 & 34  & 21,472  \\
					r/MaleForeverAlone   & 2017--2018 & 619 & 19 & 17,917    \\
					r/foreveraloneteens  & 2011--2019 & 322 & 18 & 3,592    \\
					r/Incelselfies       & 2018--2019 & 6,385 & 18  & 15,552  \\
					r/Truecels           & 2015--2016 & 364 & 17  & 5,900  \\
					r/ForeverAloneDating & 2011--2019 & 76,976 & 11 & 605   \\
					r/askanincel         & 2018--2019 & 2,465 & 10 & 2,056   \\
					r/IncelDense         & 2018--2019 & 254 & 7  & 3,410   \\
					r/SupportCel         & 2017--2019 & 352 & 6  & 2,253  \\
					\bottomrule
				\end{tabular}%
			}		
		\end{center}
		\reduce\caption{Number of videos retrieved from each data source.}
		\label{subreddit_list}
	\end{table}
	
	\begin{figure*}[t]
		\centering
		\subfigure[4chan]{\includegraphics[width=0.322\textwidth]{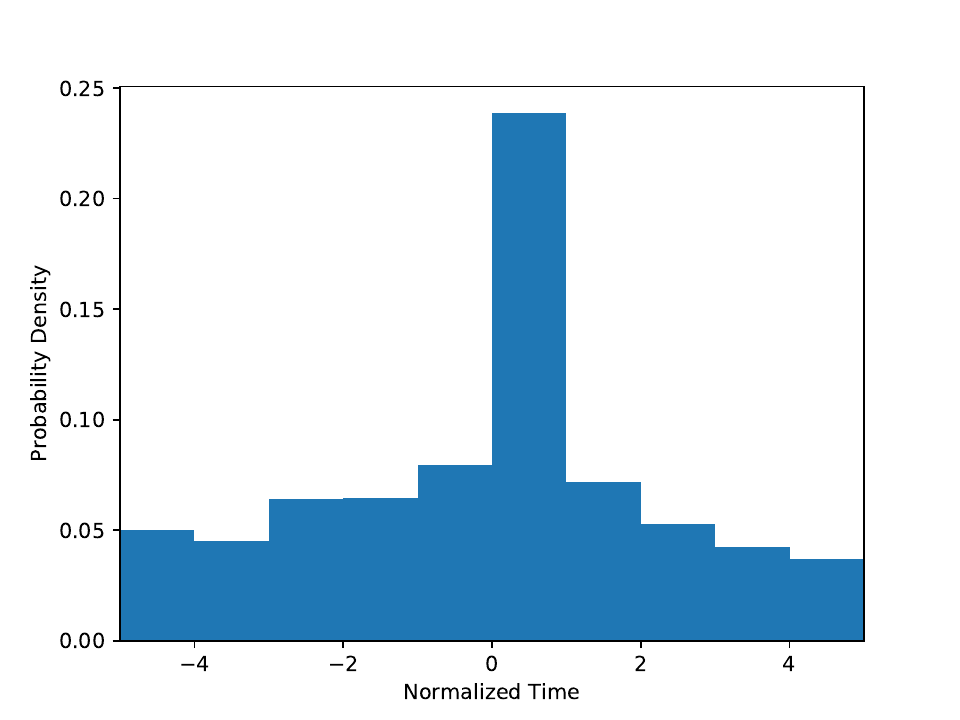}\label{fig:4chan_raid}}
		\subfigure[\donald]{\includegraphics[width=0.321\textwidth]{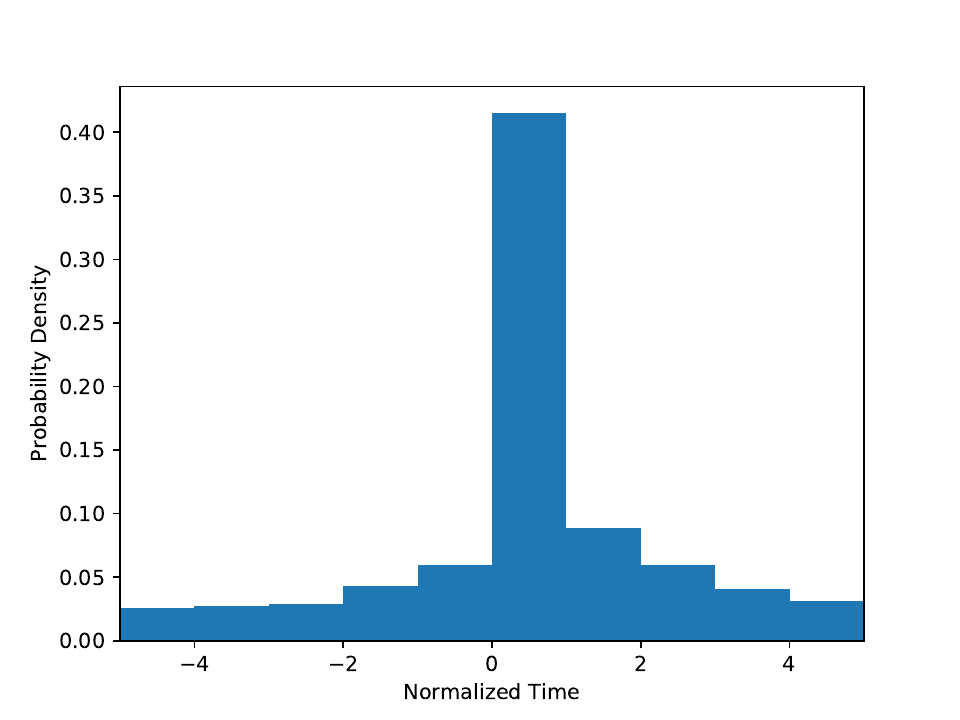}\label{fig:donald_raid}}
		\subfigure[Incels subreddits]{\includegraphics[width=0.32\textwidth]{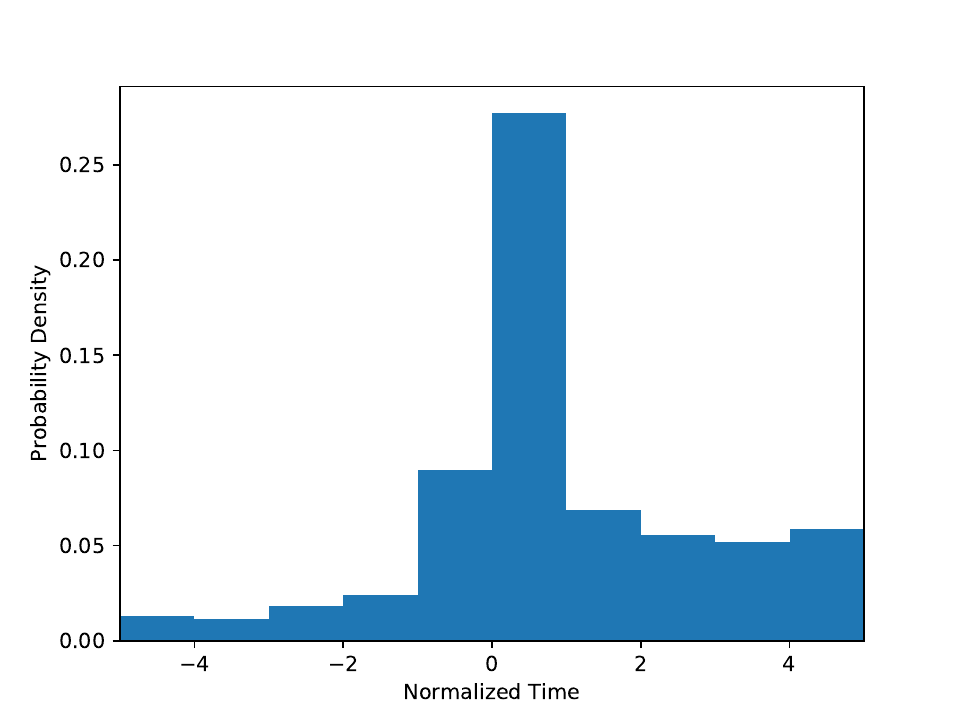}\label{fig:incel_raid}}
		\caption{Probability Density Functions (PDFs) of the activity peak in YouTube comments and the source community thread where the YouTube video is linked from. The time is normalized to the thread's lifetime, where t = 0 denotes the time when the video was first mentioned, and t = 1 is the last post in the thread.}
		\label{fig:raid} %
	\end{figure*}
	
	\subsection{Data Collection}\label{sec:data}
	Our methodology involves two steps: 1) retrieving all links to YouTube videos posted on \pol, \donald, and the 16 Incels subreddits, and 2) gathering all comments and replies to these YouTube videos.
	For \pol, we use the dataset released by~\citet{papasavva2020raiders}, which contains 3,397,911 threads and 134,529,233 posts posted between June 2016 to November 2019. 
	For Reddit, we first gather all public data posted on Reddit from 2005 to 2020, which includes 600M posts and 5B comments from 2.8M subreddits, using the Pushshift dumps~\cite{baumgartner2020pushshift}.
	Then, we filter out all comments and posts made on \donald and the 16 Incels subreddits.

	We extract all links to YouTube videos posted on these communities using regular expressions.\footnote{We find 5 different kinds of YouTube links: 1) \url{youtube.com/watch?v=video\_id}, 2) \url{youtu.be/video\_id}, 3) \url{m.youtube.com/watch?v=video\_id}, 4) \url{m.youtu.be/video\_id}, 5) \url{youtube.com/embed/video\_id}, which are captured by the regular expressions.}
	For \pol, we use the `com' field of the JSON object containing all posts from a single thread, while, for Reddit, the `url' field from the JSON object corresponding to a post.
	The total number of videos posted on 4chan is 850,523, 278,849 on \donald, and 14,616 on the Incels subreddits. 
	
	Finally, we use the YouTube Data API~\cite{youtubedatapi} to collect all comments and replies.
	We obtain 1B YouTube comments from \pol links, 63.7M from \donald, and 24.4M from the Incels subreddits. 
	An overview of our dataset is available in Table~\ref{subreddit_list}.

	\section{Characterizing Coordinated Attacks}\label{sec:motiv}
	Before we can build machine learning models for coordinated hate attacks attribution, we need to first of all ``understand'' them.
	We start by formulating two hypotheses, namely, that coordinated online attacks will share two distinctive characteristics: 1) once a YouTube link is shared on a platform, there might be a spike in its commenting activity on YouTube, which is a possible indication of a coordinated attack occurring; and 2) each community has its lingo and slang words (e.g., \pol's slang includes characteristic terms like ``cuck'' and ``libtard''), which can be modeled and used as features in a machine learning classifier.
	If these assumptions hold, coordinated hate attacks could be traced back to a source community by looking for spikes in comments and matching the language of the community with the video comments.

	To test the validity of our hypotheses, we run an experiment with videos posted on \pol, \donald, and the 16 Incels subreddits, between January and June 2019 (when \donald got quarantined). %
	We use a dataset consisting of 17,023 videos from 4chan's \pol board, 2,499 from \donald, and 248 from the 16 Incels subreddits.
	We pick videos that were only linked in one community to avoid overlaps (i.e., to have a single ground truth).
	We do this to exclude confounding variables and treat each community as a distinct class.
	This approach allows us to establish clear relationships between each source community and their possible attacks.
	
	\subsection{Commenting Activity}\label{sec:commact} 
	
	We first want to measure whether after a URL to a third platform (e.g., a YouTube video) is posted on an online community (e.g., 4chan's \pol), this third-party content observes a peak in commenting activity.
	Preliminary observations in this direction were made by Hine et al.~\cite{hine2017kek}.
	If a video linked to on a Web community receives a surge of comments during the lifetime of the discussion thread, this might indicate that a coordinated hate attack taking place.

	More formally, let $x$ be the thread on which the link to the YouTube video is posted, and $y$ the set of comments on the video. 
	We denote the timestamps in $x$ and $y$ as $\{t_{x}^{i} \mid 1,...N_{x} \}$ and $\{t_{y}^{j} \mid 1,...N_{y} \}$. %
	We then normalize the time frame for $\{t_{x}^{i} \}$ and $\{t_{y}^{j} \}$ so that $t=0$ represents the time the YouTube link is posted on the source community, and $t=1$ the time the last post is made on that thread.
	The normalized time frame is then calculated as:
	$	t =  \frac{t - t_{yt}}{t_{last} - t_{yt}} $.

	In Figure~\ref{fig:raid}, we plot the Probability Density Function (PDF) of the (normalized) YouTube comment timestamps for each of the three communities.
	The distribution shows the YouTube commenting activity with respect to the thread lifespan from where the video is linked.
	For each community, the highest peak in comment activity occurs between $t=0$ and $t=1$, which is a possible indication that an attack occurs once the link is posted on a community.
	This aligns with the results in~\citet{hine2017kek}, who find that a rise in commenting activity of a YouTube video is a possible indication of a raid taking place.

	\subsection{Language}\label{sec:language} 
	
	Our second hypothesis is that, since online communities are characterized by their own jargon, the comments originating from a certain community will present linguistic features that are closer to that community.
	To validate this assumption, we first pre-compute the TF-IDF (Term Frequency -- Inverse Document Frequency) scores of all the words in the three source communities. 
	We calculate the TF-IDF score for each word by computing TF on the given community and the IDF on the other two communities. 
	This method ensures that the score for any word used in a community shows its importance relative to the other two communities.
	For each video, we extract comments between $t=0$ and $t=1$ and calculate a TF-IDF score for each word. 
	
	Next, to understand if the comments in the video are closest to their source community, we use the keywords with the highest TF-IDF score.
	I.e., we calculate the average score of the top keywords, and for each of the three source communities, we compute the average score of the same words, finding the community with the closest language. 

	We experiment with various thresholds to identify the ideal number of keywords, ranging from Top-10 to Top-24.
	For this experiment, we use videos linked from \pol because we expect it to be the most complex and varied in terms of language.
	In Table~\ref{tbl:keywords}, we report the fraction of videos posted on \pol for which the language of the comments in the peak is closest to the one by the community itself, using various threshold selections.
	We stop at Top-24 keywords because the accuracy does not increase any further, and find that Top-20 yields the highest accuracy.

	After fixing the keyword threshold to 20, we run this approach on all videos in our dataset, reporting our results in Table~\ref{attribution}; while 85\% of the videos posted on Incel subreddits have a language that is closest to the source community, there is definitely room for improvement on \donald (45\%) and \pol (25\%).

	\begin{table}[t]
		\begin{center}
			\setlength{\tabcolsep}{3.5pt}
			\small
			\begin{tabular}{llll}
				\toprule
				\textbf{Keywords}  &  \textbf{Accuracy} & \textbf{Keywords}  &  \textbf{Accuracy}  \\ \midrule
				Top-10     &   278/1,176 (23.6\%) &
				Top-12      &  290/1,176 (24.7\%)      \\       
				Top-14     &  292/1,176 (24.8\%) &
				Top-16      &  293/1,176 (24.9\%)      \\
				Top-18      &  287/1,176 (24.4\%) &
				Top-20     &  294/1,176 (25.0\%) \\
				Top-22     &  287/1,176 (24.4\%) &				
				Top-24     &  278/1,176 (23.6\%) \\								
				\bottomrule
			\end{tabular}%
		\end{center}
		\reduce\caption{Percentage of videos posted on \pol for which the language in their comments is closest with the source community, with different keyword combinations.} %
		\label{tbl:keywords}
	\end{table}
	
	\descr{Model selection.}
	We also investigate whether more sophisticated language learning models would lead to better results.
		More precisely, we experiment with BERT (Bidirectional Encoder Representations from Transformers)~\cite{devlin2019bert}, since it is considered the state-of-the-art model for a wide range of language processing tasks like text classification~\cite{devlin2019bert}, named entity recognition~\cite{lample2019cross}, %
		text summarization~\cite{yang2019bert}, etc. %
		We use PyTorch's SentenceTransformer library to load a pre-trained Sentence-BERT (SBERT)~\cite{reimer2019sbert} model, which %
		uses siamese and triplet network structures to produce sentence embeddings that can be compared using cosine-similarity.
		We encode the text from each source community into vector representations. 
		Next, we take the videos discussed earlier in this section and encode their comments into a vector representation.
		Then, we compute the cosine similarity between the encoded representations of YouTube comments and each source community.
		Finally, we determine the most similar community based on the highest cosine similarity score.
	
	Table~\ref{attribution} shows that, cumulatively, the TF-IDF-based approach outperforms BERT, correctly attributing 790 out of 2,242 videos, %
		while the BERT-based model correctly attributes 756.
		In particular, on \donald, it only attributes 97 out of the 985 videos correctly. 
		Therefore, we select TF-IDF due to its simplicity, consistency, lower resource overhead, and better performance.

	\subsection{Controlling for Lag} 
	To improve accuracy, we opt to control for the lag and the number of comments on the YouTube video. 
	As hypothesized by~\cite{hine2017kek}, in the case of an attack, the comments of the source community and the YouTube video are likely to be synchronized; hence, the lag is close to 0.
	To estimate the lag between the two signals (i.e., the posts on the source community and the comments on the YouTube video), we use cross-correlation and run a grid search to identify the best lag range and the number of comments on the YouTube video.
	
	For \pol, we get the highest accuracy (71\%) with a comment range of [42-72] and a lag of 0.
	For \donald and the Incels subreddits, the best settings are, respectively, with a comment range of [65-100] and a lag range of [0-1], and a comment range of [40-100] without any lag restriction, yielding, respectively, 85\% and 100\% accuracy. 
	
	Overall, our analysis shows that once a YouTube link is shared on a polarized community, there is a spike in comments on the video, and the language of these comments is closest to the source community. %
	We use these observations to design our raid attribution system, \system.

	\begin{figure*}[!t]
		\centering
		\includegraphics[width=0.95\textwidth]{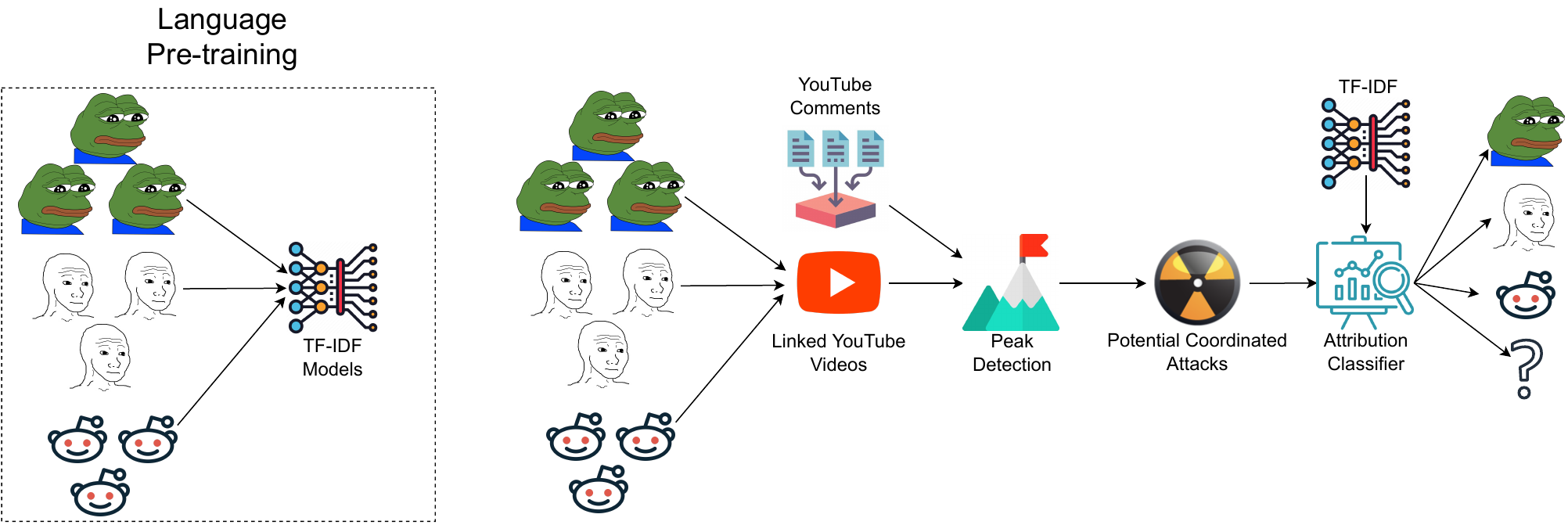}
		\caption{Overview of \system: a set of communities are fed to the system, in this case: (1) 4chan's \pol board, (2) r/The\_Donald subreddit, and (3) 16 Incels subreddits. \system learns their language through TF-IDF on the top keywords. To attribute potential attacks, it collects all YouTube comments on videos linked from each source community and identifies peaks in the comment activity of these videos as an indication of a potential coordinated attack. Finally, \system attributes attacks back to a source community using a machine learning classifier based on the TF-IDF scores of top keywords.}	\label{fig:systempipeline}
	\end{figure*}

	\begin{table}[t]
		\begin{center}
			\small{
				\begin{tabular}{lrr}
					\toprule
					\textbf{Community}  &  \textbf{TF-IDF} & \textbf{S-BERT}  \\ \midrule
					\pol     &  \textbf{294/1,176 (25\%)} & 624/1,176 (53\%) \\
					\donald     &     \textbf{427/985 (43\%)}  & 97/985 (10\%) \\       
					Incels subreddits     &  \textbf{69/81 (85\%)} & 35/81 (43\%) \\
					\bottomrule
				\end{tabular}%
			}
		\end{center}
		\reduce\caption{Accuracy of attributing attacks on videos linked from each source community.}
		\label{attribution}%
	\end{table}

	\section{\system}
	
	In this section, we present the three main components of the \system system, namely, 1) language pre-training, 2) peak detection, and 3) attribution.
	\system's pipeline is also depicted in Figure~\ref{fig:systempipeline}: \system first learns the language used in all the source communities presented to it; then, for each YouTube video shared in these communities, it detects peaks in their commenting activity.
	Finally, it uses a machine learning classifier to attribute the comments during the peak to one of the source communities.
	
	\system only requires two sets of timestamped messages, one from the source community (e.g., a \pol thread) and one from the target community (e.g., comments on a YouTube video). 
	Therefore, our approach could be easily adapted to other services as well .%

	\subsection{Language Pre-Training}\label{sec:pretrain}
	To model the language used by different communities when carrying out aggression attacks, we use the methodology discussed in Section~\ref{sec:language} and perform a TF-IDF analysis to identify the words that stand out the most from the attacks by a certain community.
	
	We pre-compute TF-IDF scores of all words in the three source communities. 
	We calculate the TF-IDF score for each word by computing TF on the given community and the IDF on the other two communities. 
	
	\subsection{Peak Detection}\label{sec:peak}
	Next, we detect peaks in the commenting activity of a video, which we can later attribute to a source community based on linguistic features.
	From Section~\ref{sec:commact}, we know that there is a surge in YouTube video comments once a link to it is posted on a community.
	Therefore, \system relies on a peak detection module to identify deviations in comment activity that can be tested for potential coordinated attacks.
	
	To identify peaks, we calculate the daily mean number of comments to the video, along with the standard deviation, for all the days we have comments for a specific video.
	Whenever \system encounters a day or period with more comments than the sum of daily mean and standard deviation, it labels it as {\em a period with a peak} in the commenting activity of a video, and thus worth to be examined.

	\subsection{Attribution}
	Once a peak is identified, we extract all comments during the relevant time range.
	\system trains a multi-class classifier to attribute a given attack to the source community.
	Using the TF-IDF scores obtained from Section~\ref{sec:pretrain}, we extract the Top-20 keywords from each source community.
	The TF-IDF scores for the selected words serve as the features of a given video.
	In other words, each video has 60 features (i.e., 20 words from each community).
	
	Then, we calculate a TF-IDF score for each word using the comments during the peak of each YouTube video.
	For words that do not appear in the comments of the video, their feature value is set to 0.
	Based on the TF-IDF scores, \system trains a supervised model to identify the community which initiated the attack on the given YouTube video. 
	The model trains on a set of labeled videos, where the label is a source community.
	On an unseen video, \system outputs one of four labels, corresponding to Incels subreddit, \donald, \pol, or that the video cannot be attributed to any of the three source community.

	Overall, our attribution classifier allows us to determine with high confidence the community that launched a raid without looking at the context of the text being discussed in the comment thread.
		We believe this is preferable to simply checking whether or not a link has been shared on a platform; it is also less prone to content drift since, while the topics being discussed might change, we expect the language used by entire communities to change at a slower pace. %

	\section{Evaluation}\label{sec:eval}
	In this section, we present the results of our experimental evaluation for each component of \system's pipeline.
	
	\begin{table}[t]
		\begin{center}
			\setlength{\tabcolsep}{3pt}
			\small
			\resizebox{\columnwidth}{!}{%
				\begin{tabular}{rrrr|rrrr}
					\toprule
					& 				\textbf{\pol}  &  \textbf{T\_D} & \multicolumn{1}{r}{\textbf{Incels}} 
					&& 				\textbf{\pol}  &  \textbf{T\_D} & \textbf{Incels}\\ \midrule
					1&				fuck & trump    &   incel & 11&				know & cnn &  make \\
					2&				nigger &peopl    &   women& 12&				kike &fake&  thing \\
					3&				don &think     &  men & 13&				now &need&  date \\
					4&				white &news    &   girl & 14&				faggot &right&  life \\
					5&				peopl &make    &   guy & 15&				us &fuck&  chad \\
					6&				jew &vote     &   one & 16&				time &hillari&  realli \\
					7&				shit &know&  sex & 17&				right &even&  friend \\
					8&				make &time&  feel & 18&				good &presid&  good \\
					9&				want& want & person & 19&				countri &thing&  someon \\
					10&				think &content&  attract & 20&				trump	& clinton&  relationship\\						
					\bottomrule
				\end{tabular}%
			}
		\end{center}
		\reduce\caption{Top-20 keywords with the highest TF-IDF scores for each source community. (T\_D denotes \donald).}
		\label{tbl:commkeywords}%
	\end{table}

	\subsection{Pre-Training}
	As discussed, we perform a TF-IDF analysis on all three source communities.
	We find a total of 1,610,197 unique words on \pol, 401,738 on \donald, and 144,252 unique words on the Incels subreddits after removing stop words and using Porter Stemmer~\cite{porter_stemmer} to find the word stems.
	This difference in the number of words used in each community is a relatively strong indication of the difference in the language used in each community.
	
	Table~\ref{tbl:commkeywords} shows the Top-20 keywords for each source community; \donald includes more political keywords, \pol a lot of racist and abusive words, and the Incel subreddits more dating-related keywords.
	
	\subsection{Peak Detection}\label{sec:peakeval}
	We detect peaks in YouTube videos as per the methodology in Section~\ref{sec:peak}.
	As illustrated in Figure~\ref{fig:systempipeline}, \system feeds the comments from the peak to the attribution module.
	Depending on the length and comments of the video, a number of peaks may be identified in any given video.
	
	From the 20,325 videos linked from all communities in the first six months of 2019, we find that 562 videos do not have any peaks.
	The other 19,763 videos have at least a peak---more precisely, 17.5 peaks per video on average.
	To optimize the attribution, \system only considers as optimal peaks that have a certain number of comments, which we denote as ``Minimum Comments Threshold.''
	
\begin{figure}
		\centering
		\includegraphics[width=0.45\textwidth]{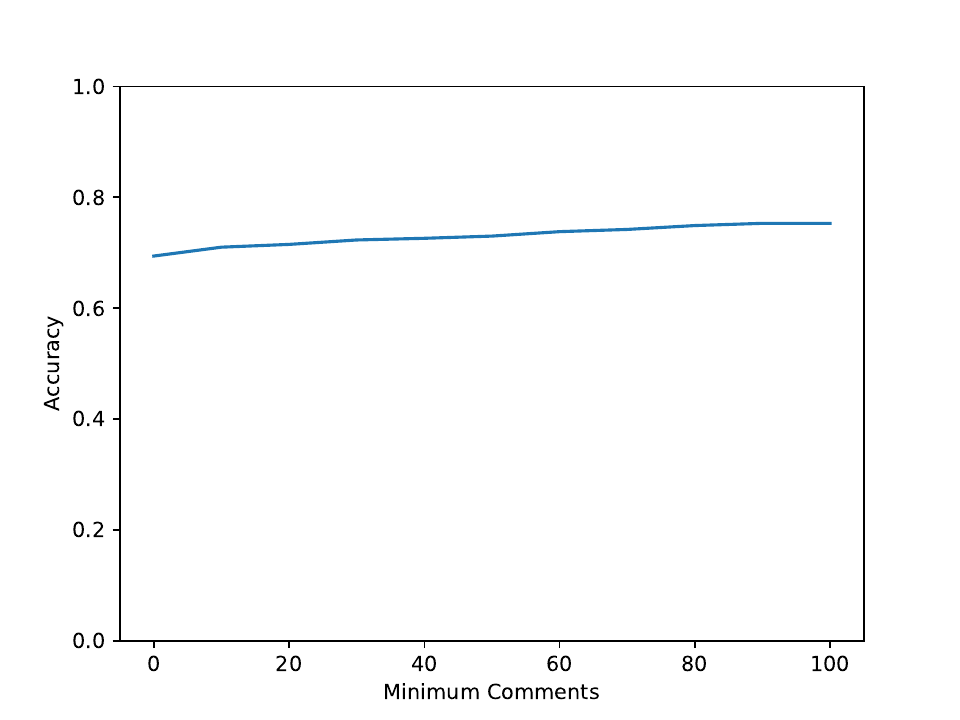}
		\caption{Attribution accuracy for various "Minimum Comments" thresholds.}
		\label{fig:peakacc}%
	\end{figure}

	We believe this threshold is best treated as a hyper-parameter and thus tune it as part of the experiments presented in Section~\ref{sec:attrib}, aiming to find the optimal value which produces the best classification results (see Figure~\ref{fig:peakacc}).

	\subsection{Attribution}\label{sec:attrib}
	To create our attribution pipeline, we need a set of videos that have not been linked from one of the three source communities.
	Therefore, we build a dataset of 50 unrelated YouTube videos. 
	We use the YouTube Data API search feature to query videos for the following keywords: sports, gaming, and music.

	To train our system, we curate a set of 200 total videos, 150 of which are linked from either of the three source communities, selected at random: 1) \pol, 2) \donald, and 3) the Incels subreddits.
	These 150 videos are randomly sampled from our dataset from the first six months of 2019 (before \donald got quarantined).
	The other 50 is the set of unrelated YouTube videos, i.e., not linked from our source communities.
	We perform ten-fold cross-validation and evaluate the performance of four classifiers based on accuracy, precision, recall, and F1 score; please see
	Table~\ref{classificationResults}.
	We find that Random Forest yields the best performance, with $75.3\%$ accuracy in correctly attributing a video to its source community.
	
	As mentioned in Section~\ref{sec:peakeval}, we treat the number of comments required to be considered a peak as a hyper-parameter.
	We experiment with various thresholds by linearly increasing the minimum number of comments by ten during training and analyzing its impact on attribution accuracy.
	The resulting accuracy for each minimum comment threshold is reported in Figure~\ref{fig:peakacc}: there is a steady increase in accuracy until 90 comments, and then it levels off.
	As a result, we set the minimum comments threshold to 90, which gives us a total of 221 peaks for 200 videos.
	
	\descr{Comment Threshold Analysis.}\label{sec:commentthreshold}
	To ensure that enforcing a minimum comment threshold does not prevent \system from attributing real-world aggression attacks, we manually analyzed a set of videos with under 90 comments to attribute potential aggression attacks.
	We picked 33 videos linked from each source community and had two annotators independently assess the comments for hateful, aggressive, toxic, and inflammatory speech markers.
	
	Annotator 1 found two potential aggression attacks, while Annotator 2 found one, in common with Annotator 1.
	This yields a Cohen's Kappa score~\cite{cohen2022} of 0.662 (high agreement).
	As a result, we estimate an upper-bound error rate of approximately 2\% (Annotator 1 found 2/99 videos to be potential aggression attacks).
	This confirms the validity of considering the threshold as a hyper-parameter that can be fine tuned as necessary.
	
	Next, we compute a confusion matrix to understand how well our system labels videos of each class, in particular, the unrelated videos.
	Class 1 corresponds to the Incels subreddits, 2 to \donald, 3 to \pol, and 4 to the unrelated videos.
	As shown in Table~\ref{confMat}, our system can correctly identify the unrelated videos as well.

	\begin{table}[t!]
		\begin{center}
			\setlength{\tabcolsep}{5pt}
			\small{
				\begin{tabular}{lrrrr} 
					\toprule
					{\bf Classifier} & {\bf Accuracy} & {\bf Precision} & {\bf Recall} & {\bf F1-Score}  \\ 
					\midrule
					Linear SVM & 34.0\% & 10.0\%	& 29.2\%	& 19.6\% \\
					Decision Tree & 62.8\% & 49.3\% 	& 47.1\%	& 48.2\% \\ 
					KNN	& 63.5\% & 54.9\%	& 55.5\%	& 55.2\%  \\ 
					Random Forest& \textbf{75.3\%} & \textbf{70.5\%}	&	\textbf{67.8\%}	& \textbf{69.2\%} \\
					\bottomrule
				\end{tabular}
			}
		\end{center}
		\reduce\caption{Performance of the different classifiers using ten-fold cross validation.}
		\label{classificationResults}%
	\end{table}

	\subsection{Misclassified Videos}\label{sec:misclass}
	As is common with classification tasks, our attribution module misclassifies a (relatively small) fraction of the videos.
	Using the confusion matrix in Table~\ref{confMat}, we observe that the False Positive Rate (FPR) for the Incels subreddits is 0.033 and 0.067 for the other three classes.
	The highest percentage of false negatives is on \pol, with a False Negative Rate (FNR) of 0.4; all the other classes have an FNR of 0.1.
	
	In the rest of this section, we shed light on the misclassifications and present a few relevant examples.  %
	To make things easier, we divide the misclassified videos into two categories: 1) Crossover Videos and 2) No Aggression Videos.
	
	\descr{Crossover Videos.}\label{sec:cross}
	A crossover video features concepts more relevant to another community than the one it was linked from, e.g., a video discussing President Donald Trump on an Incels subreddit.
	A very small minority of these videos, especially those showing heavy overlap with another community, are misclassified because the comments dilute the specificity of the language of the original source community.
	For instance, the video ``Mick Mulvaney on Trump's booming economy'' is linked from 4chan, but being related to politics and President Donald Trump, the commenting activity uses a language that is more related to those topics and more similar to \donald' lingo.
	
	This is arguably a limitation inherent to language modeling.
	\system uses TF-IDF values as features for attribution; future work could investigate the use of different language model tools to be less susceptible to this phenomenon.

	\begin{table}[t!]
		\begin{center}
			\setlength{\tabcolsep}{4pt}
			\small{
				\begin{tabular}{lrrrr} 
					\toprule
					{\bf Predicted\textbackslash Actual} & {Incels} & {\donald} & {\pol} & {Unrelated} \\
					\midrule
					{Incels}	&	9	&	0	&	0  & 1 \\
					{\donald} &  0    &  9     &  1    & 0 \\ 
					{\pol}	& 1	&	2	&	6 & 1  \\ 
					{Unrelated}	& 0	&	0	&	1  & 9 \\
					\bottomrule
				\end{tabular}
			}
		\end{center}
		\reduce\caption{Confusion Matrix of our classification model.}
		\label{confMat}%
	\end{table}

	\descr{No Aggression Videos.}
	The second kind of misclassified videos are those not attacked by the source community.
	A large majority of them are music videos with many comments and views, linked from platforms like 4chan and Reddit simply for entertainment purposes, often not receiving any traction in the community.
	A relevant example is ``Earth, Wind \& Fire -- Let's Groove,'' which has 49.4K comments on YouTube and 231M views as of October 4, 2022.
	Although it has been linked from 4chan, \system labels it as ``unrelated'' or, in other words, not attacked by any community, which means that the language in the detected peak(s) does not match the source community.
	Hence, we postulate that the peaks in videos shared for non-attack purposes are largely observed due to other factors, e.g., artists sharing the music video on their social media, a paid traffic boost, or promotions.
	However, just because it is coincidentally linked from one of the platforms does not imply that an actual attack is taking place.
	As a side note, in this case, the original thread on 4chan only received one comment.
	Therefore, it is safe to assume that in this case and other similar cases, there was no actual aggression attack.
	
	\begin{table}[t!]
		\begin{center}
			\setlength{\tabcolsep}{0.5pt}
			\resizebox{\columnwidth}{!}{%
				\begin{tabular}{lrrrrrrr} 
					\toprule
					{\bf Perspective} & \small {Attr.} &  \small {NonAttr.} & \small {Baseline} & \small {Ks 1} & \small {P-Value} &\small  {Ks 2} &  \small {P-Value}\\
					\midrule
					{Toxicity}	&	0.246	&	0.218	&	0.147  & ~0.069 & \textbf{~$<$\pval} & ~0.199 & \textbf{~$<$\pval}\\
					{Severe Toxicity} &  0.035    &  0.032     &  0.015    & 0.040 & \textbf{$<$\pval} & 0.171 & \textbf{$<$\pval} \\ 
					{Identity Attack}	& 0.071	&	0.059	&	0.034 & 0.071 & \textbf{$<$\pval} & 0.231 & \textbf{$<$\pval}  \\ 
					{Insult}	& 0.166	&	0.133	&	0.092 & 0.085 & \textbf{$<$\pval} & 0.215 & \textbf{$<$\pval}  \\ 
					{Profanity}	& 0.142	&	0.136	&	0.087 & 0.026 & \textbf{$<$\pval} & 0.131 & \textbf{$<$\pval}  \\ 
					{Threat}	& 0.036	&	0.036	&	0.021 & 0.021 & \textbf{$<$\pval} & 0.117 & \textbf{$<$\pval}  \\
					{Sexually Explicit}	& 0.067	&	0.064	&	0.044 & 0.065 & \textbf{$<$\pval} & 0.079 & \textbf{$<$\pval}  \\ 
					{Flirtation}	& 0.374	&	0.377	&	0.349 & 0.020 & \textbf{$<$\pval} & 0.093 & \textbf{$<$\pval}  \\ 
					{Attack on Author}	& 0.148	&	0.149	&	0.144 & 0.042 & \textbf{$<$\pval} & 0.062 & \textbf{$<$\pval}  \\ 
					{Att. on Commenter~}	& 0.267	&	0.270	&	0.222 & 0.016 & \textbf{$<$\pval} & 0.085 & \textbf{$<$\pval}  \\ 
					{Incoherent}	& 0.514	&	0.535	&	0.648 & 0.044 & \textbf{$<$\pval} & 0.248 & \textbf{$<$\pval}  \\ 
					{Inflammatory}	& 0.391	&	0.341	&	0.270 & 0.088 & \textbf{$<$\pval} & 0.225 & \textbf{$<$\pval}  \\ 
					{Likely to Reject}	& 0.674	&	0.660	&	0.606 & 0.022 & \textbf{$<$\pval} & 0.121 & \textbf{$<$\pval}  \\ 
					{Obscene}	& 0.281	&	0.272	&	0.233 & 0.014 & \textbf{$<$\pval} & 0.082 & \textbf{$<$\pval}  \\ 
					{Spam}	& 0.111	&	0.136	&	0.207 & 0.085 & \textbf{$<$\pval} & 0.254 & \textbf{$<$\pval}  \\ 
					{Unsubstantial}	& 0.640	&	0.665	&	0.711 & 0.053 & \textbf{$<$\pval} & 0.108 & \textbf{$<$\pval}  \\ 
					\midrule
					{\bf Rewire}  & \small {Attr.} &  \small {Non Attr.} & \small {Baseline} & \small {Ks 1} & \small {P-Value} &\small  {Ks 2} &  \small {P-Value}\\
					\midrule
					{Abuse}	& 0.293	&	0.241	&	0.226 & 0.061 & \textbf{$<$\pval} & 0.082 & \textbf{$<$\pval}  \\ 
					{Hate}	& 0.157	&	0.132	&	0.098 & 0.086 & \textbf{$<$\pval} & 0.138 & \textbf{$<$\pval}  \\ 
					{Sexually Explicit}	& 0.032	&	0.037	&	0.052 & 0.023 & \textbf{$<$\pval} & 0.078 & \textbf{$<$\pval}  \\ 
					{Profanity}	& 0.111	&	0.106	&	0.071 & 0.006 & 0.301 & 0.040 & 0.001  \\ 
					{Positive}	& 0.168 &	0.205	&	0.198 & 0.075 & \textbf{$<$\pval} & 0.133 & \textbf{$<$\pval}  \\ 
					{Violent}	& 0.039	&	0.034	&	0.020 & 0.025 & \textbf{$<$\pval} & 0.105 & \textbf{$<$\pval}  \\ 
					\bottomrule
				\end{tabular}
			}
		\end{center}
		\reduce\caption{Toxicity scores for each video category.}
		\label{toxScores}%
	\end{table}

	\subsection{Detection in the Wild}\label{sec:wild}
	Finally, we test \system in the wild to attribute attacks to a given source community. 
	We use the data collected in Section~\ref{sec:data} and extract 3,333 videos per community, obtaining a total of 9,999 videos.
	We use this data as we have ground truth; i.e., we know which community a given video was linked from. 
	\system attributes 700 videos out of the 9,999 to the labeled source community, while 8,644 videos did not pass our minimum peak threshold of 90 comments and were discarded.
	
	We manually review the attributed videos to understand the attacks; in Section~\ref{sec:case_studies}, we discuss a few examples of what these aggression attacks look like, from the original thread to the YouTube video comments.

	\subsection{Raid vs Commenting Activity}

	One intuition behind \system is that \textit{peaks in commenting activity} on YouTube videos, combined with the \textit{use of language similar to that of hateful online communities}, strongly suggests the occurrence of raids. 
	Both factors play a crucial role in this determination.
	To further validate this intuition, it is important to differentiate between benign commenting activity that can appear as coordinated and an actual raid.
	The latter involves real, targeted hate; the former means that the video is merely receiving attention of some kind (e.g., when a video link is shared across platforms without any malicious intent).
	Thus, we set out to analyze the commenting activity in videos attributed by \system and compare them to non-raided videos using various metrics that are indicative of raids (e.g., toxicity, abuse, hate, and inflammatory content).
	
	We divide the videos into: 1) attributed videos or raids identified by \system in Section~\ref{sec:wild}, 2) non-attributed videos from Section~\ref{sec:wild} that had peaks and passed the minimum comment threshold but were not classified as raids, and 3) 50 generic videos of various categories (e.g., music and gaming) from Section~\ref{sec:attrib} that serve as our baseline.
	For each video, we randomly pick 50 comments from their peaks and feed them to Google's Perspective API~\cite{perspectiveAPI} and the Rewire API~\cite{rewire2023}.\footnote{Perspective API is an open-source tool that assigns [0-1] scores to a given text across several metrics. Rewire is another machine learning-based tool that detects toxicity in text, developed in partnership with Nvidia, also assigning [0-1] confidence scores.
	}
	Our intuition is that if the commenting peaks attributed by \system as raids show significantly higher values for those metrics, then this is a strong indication that they are raids.

	The average scores for each metric from both APIs and the related video categories are reported in Table~\ref{toxScores}.
	Our baseline sample reveals that even generic videos contain some degree of hate speech and negativity.
	Due to the controversial nature of videos published on platforms like 4chan, we observe a noticeable increase in inflammatory markers for both attributed and unattributed videos.
	However, raided videos display the highest levels of toxicity, hate, and abuse, even though \system does not consider toxicity scores when performing attribution.
	
	\descr{Significance Testing.} We perform the Kolmogorov–Smirnov~\cite{kstest2022} test to determine whether the differences in scores are statistically significant.
	For each metric, we perform two tests: 1) between attributed and non-attributed videos, denoted as Ks-Score 1, and 2) between attributed and baseline videos, denoted as Ks-Score 2.
	We report these scores in Table~\ref{toxScores}, along with P-values.
	As we are testing multiple hypotheses at once, which increases the chance of rejecting the null hypothesis by chance, we use Bonferroni Correction~\cite{bonferroni2021}, where we adjust the significance level alpha ($\alpha$).
	We set the initial value of $\alpha$ to 0.01 and divide it by the number of hypotheses we are testing.
	We test 44 hypotheses (22 metrics and 2 hypotheses per metric), therefore the adjusted value of $\alpha$ is \pval. 
	As a result, we only reject the null hypothesis if the P-value is $<$\pval.
	Overall, we can reject the null hypothesis for all metrics except ``profanity,'' i.e., the differences in all the other samples are statistically significant. %
	
	\descr{Implications.} Note that Rewire defines abuse as ``content that is insulting, aggressive, or threatening.''
	It defines hate as ``abuse targeted at a protected group or at its members for being a part of that group, where protected groups are based on characteristics such as gender, race, or religion'' and profanity as `` a word or expression that is socially or culturally offensive, usually due to being obscene or explicit.''
	We postulate that \system is not a simple offensive speech detection tool, but it can also be used to attribute based on targeted abuse and hate, as shown by the difference in scores for the respective metrics and their associated significance. 
	
	\section{Case Studies}\label{sec:case_studies}
	To shed further light on coordinated attacks attributed by \system, we now discuss a few selected case studies.
	We do so aiming to analyze the modus operandi of the actors participating in the attacks with respect to spreading hate and how they are directed by the communities.
	
	A common thread in these attacks is the aggressors often posting a controversial link that is relevant to the source community to direct hate and challenge the idea presented in the video. 
	The examples we provide next show how posts in all three communities lead to aggression attacks that range from political to social issues.

	\subsection{\donald}
	One relevant example from \donald relates to the submission ''Never Go Full Feinstein: DiFi Plans on Introducing New Anti-2A Bill (banning 18-20 year olds from purchasing any gun, Unconstitutional).''
	The submission has 1,351 upvotes and 63 comments.
	Roughly speaking, \donald sees a lot of activity from the alt-right community, which is usually vocal against any restrictions to gun ownership, and thus against this bill.
	
	Here are some comments from the Reddit thread:
	
	\begin{mdframed}[style=exampledefault, backgroundcolor=vlightgray]
		\small
		\textbf{COMMENT 1:}
		I hate that fuckin witch.
		
		\noindent \textbf{COMMENT 2:}
		She was born in the 1920's and is still alive. She's probably being given baby blood transfusions or something.
		
		\noindent \textbf{COMMENT 3:}
		So your old enough to lay your life on the line in the military to defend these fucks but not old enough to own a gun for you or your families protection...I see...
	\end{mdframed}
	
	The first two are directed to Senator Feinstein, and one of those about her being a ``witch'' was also later found on the YouTube video as shown in the next snippet.
	The comments below are a short snippet of comments made on YouTube during the time this thread was active on Reddit.
	We remove any information that identifies the posters in the YouTube comments. %

	\begin{mdframed}[style=exampledefault, backgroundcolor=vlightgray]
		\small
		\textbf{COMMENT 1:}
		Wicked witch of the west
		
		\noindent \textbf{COMMENT 2:}
		We have to do something, even if it doesn't work. So people think that we are doing something.
		
		\noindent \textbf{COMMENT 3:}
		She can introduce all the anti-2A bills she wants. I will defy and ignore every single one.
	\end{mdframed}

	The discussion on Reddit was largely against the bill and Feinstein.
	We observe the same trend in comments around that time period on the YouTube video.
	By looking at the comments on the submission and the snippet from the comments on the YouTube video, we can better understand how these communities push their ideas and spread hate speech at the same time. 
	The comments on the YouTube video, like those on Reddit, are also against the bill and Sen.~Feinstein.
	
	\subsection{Incels}
	Next, we discuss a post made in the ``IncelTears'' subreddit.
	The post includes a YouTube video link to an old music video from the Donna Reed Show.
	The video is shot in a traditional setting with a woman singing songs and wearing a sundress typical of the 1950s and 60s.
	The post on Reddit, as shown below, paints the scene in two ways: (1) how women show much more skin in their dressing and are less feminine in today's day and age, and (2) women date men based on ``hypergamy,'' in other words, they date men of a higher status than themselves while rejecting lower value males.
	
	\begin{mdframed}[style=exampledefault, backgroundcolor=vlightgray]
		\small
		It's over for Incels: Incels love to talk about feminism causing `hypergamy,' but this scene from the pre-second-wave-feminist Donna Reed Show, set in an wholesome incel fantasy world, features a song that is literally about a girl who rejects all other boys because she only wants Chad
	\end{mdframed}
	
	The comments on Reddit, a few of which are shown below, carry the same narrative as the post.
	The whole idea is that women were far prettier and more feminine back in the day, as shown in comments 2 and 3.
	However, the notion that women's dating preferences are based on ``hypergamy'' is highlighted in the first comment, where the user complains about his mom rejecting low-value males the same way all females do.
	
	\begin{mdframed}[style=exampledefault, backgroundcolor=vlightgray]
		\small
		\textbf{COMMENT 1:}
		Yeah that's like the least problematic thing my mom did by far.
		
		\noindent \textbf{COMMENT 2:}
		Women back then were prettier, in my humble opinion.
		
		\noindent \textbf{COMMENT 3:}
		Feminism should have been stopped at the first wave.
	\end{mdframed}
	
	The comments on YouTube from the same time as the Reddit post are shown below.
	While the general comments on the YouTube video are in praise of the singer and the TV show in general.
	The comments from the time of the Reddit post carry the same narrative of women becoming more masculine.
	Comment 1 and 2 clearly show the users being enraged that women ``belong in the kitchen'' and that they used to be feminine back in the day.
	The last comment goes to show that obscene language is often a characteristic of these aggression attacks. 
	
	\begin{mdframed}[style=exampledefault, backgroundcolor=vlightgray]
		\small
		\textbf{COMMENT 1:}
		Why would she go to college? You don't need a college education to clean house and cook.
		
		\noindent \textbf{COMMENT 2:}
		Such a beautiful girl, and a voice to match..Gives me goose bumps to think that girls back then  were feminin..No ugly tattoes, no fking and blinding, trying to act like men etc..Take me back there..
		
		\noindent \textbf{COMMENT 3:}
		Using blown up condoms for gym decorations...even back then
	\end{mdframed}

	\subsection{\pol}
	Finally, we discuss a post on 4chan's \pol board, where the poster shares a YouTube video link of a woman performing stand-up comedy.
	The author claims that women are not funny and cannot be good comedians.
	
	\begin{mdframed}[style=exampledefault, backgroundcolor=vlightgray]
		\small
		Why the fuck are there no funny women on this planet? Why do they suck so much at comedy? Just look at this cringe
	\end{mdframed}
	
	The post received 305 replies and some of the comments are highlighted below.
	Ostensibly, the posters believe that women are not smart, are incapable of understanding and appreciating humor, and that there is no biological incentive for them to be funny.
	
	\begin{mdframed}[style=exampledefault, backgroundcolor=vlightgray]
		\small
		\textbf{COMMENT 1:}
		Women aren't mentally capable of intelligent humour, why would they?
		
		\noindent \textbf{COMMENT 2:}
		Because women really aren't very smart.
		
		\noindent \textbf{COMMENT 3:}
		They have no evolutionary/reproductive incentive for being funny.
	\end{mdframed}
	
	The comments below are taken from YouTube at the time that the video was shared on 4chan.
	The same idea is pushed in the comments that women are inherently not funny.
	
	\begin{mdframed}[style=exampledefault, backgroundcolor=vlightgray]
		\small
		\textbf{COMMENT 1:}
		Has there ever been a funny woman?
		
		\noindent \textbf{COMMENT 2:}
		female comedian
		
		\noindent \textbf{COMMENT 3:}
		That classic tactic, when the audience won't laugh at my jokes, passively aggressively bitch at them because clearly it's their fault.
	\end{mdframed}
	
	\subsection{Remarks} 
	Overall, these case studies illustrate how coordinated attacks are carried out in various ways, depending on the ideology of the attacking community.
	In certain instances, as the \donald example demonstrates, a specific person is targeted with hate speech.
	In others, the targets are broader, for instance, with misogynistic and abusive words directed at women in general.
	
	Moreover, the rhetoric used in these attacks can be exceptionally condescending and degrading, making them particularly harmful because they target actual individuals.
	
	\section{Discussion \& Conclusion}
	In this paper, we presented \system, a generalizable system to attribute coordinated aggression attacks to the community that organized and carried them out. 
	Our experimental analysis showed we can attribute attacks on YouTube videos to a source community with an accuracy of over 75\%.
	We demonstrated that coordinated attacks result in peaks in the activity of YouTube video comments and that the language used in the comments varies depending on the community the video was linked from.
	These linguistic traits enabled us to identify the communities where an attack originated.
	\system uses language features to train a machine learning classifier that attributes peaks in YouTube video comments to a source community.
	We also presented case studies of several aggression attacks identified by \system with overt elements of hate speech and misogyny to emphasize the rising issue.
	
	\subsection{Broader Perspective}\label{sec:broad}
	
	\descr{Positive Implications.}
	Effective and efficient attribution of coordinated hate attacks enables more nuanced and context-dependent moderation strategies.
	For instance, if the purpose behind an attack is believed to be racism, the platform could, in addition to deleting the hateful content, point  victim to resources relevant to dealing racism.
	Attribution can also be used in soft moderation schemes (e.g., ``this comment is very similar to those originating from 4chan''), which leverage the public's increasing understanding of the damage wrought by the darker parts of the Web.

	Overall, \system %
	can help platforms devise tailored policies.
	Previous work on personalized content moderation is limited because approaches either lack the required nuance~\cite{chandrasekharan2019crossmod} or solely rely on the manual selection of relevant keywords by content creators~\cite{jhaver2022designing}.
	\system allows going beyond that, as the language models built for each source community can be used to automatically generate blocklists of keywords for attacks orchestrated by those communities.
	
	An alternative deployment of \system could be to prioritize moderation of source communities based on how ``dangerous'' their attacks are.
	Since human vetting is the bottleneck of the moderation efforts by many online platforms, this prioritization could help quickly get to the attacks with the highest risk of causing damage to their victims.
	
	\descr{Potential Negative Outcomes.}
	Moderation techniques like \system are tantalizing in that their positive impact is pretty clear.
	At the same time, the worst-case scenario for deploying a moderation strategy is, arguagly, not that it would just not work; rather, it might make the problem \emph{worse} as deplatformed users move onto more extreme and echo-chamber-y platforms~\cite{ali2021understanding}.
	Thus, we need to consider several avenues that should be considered in more detail before \system should be deployed.
	
	Malevolent actors who wish to coordinate hate attacks might use \system to learn whether they are at ``risk'' of being attributed, and change their strategy until the system misclassifies them.
	However, raids are conducted by several independent attackers stemming from the same platform, and the level of tight coordination required to evade \system would be difficult to achieve.

	Moreover, false positives from \system are not entirely different from accusing a community of a crime they did not commit.
	This has obvious problems from a ``justice'' point of view, but it also has the potential to stymy reformation processes that might be occurring within the attacking community (whether organic or due to some other mitigation).
	While it is true that 4chan is the source of many coordinated aggression attacks, claiming that they were responsible for attacks they did take part in only serves to reinforce some of the conspiratorial narratives that motivate the attacks in the first place.
	
	Finally, we make no recommendation as to what specific moderation action(s) should be taken following the attribution of a coordinated hate attack, but at least some choices could open up additional ways to execute attacks.
	For example, if a deployment were to automatically provide a victim of an attack with resources, a naive implementation might end up allowing attackers to flood the victim with messages linking to those resources, which could itself be triggering.
	Even though this is a trivial example, and one that we would argue is still an improvement over the status quo, it does highlight that even effective solutions to the problem of coordinated hate attacks (and other content moderation tasks) have rough edges.

	\descr{Ethics Considerations.} %
	Since we do not work with human subjects and only use data available to the public, our work is not categorized as human subjects research by our institution's Institutional Review Board (IRB).
	We follow common ethical standards; we do not attempt to de-anonymize users, we remove all personally identifiable information from the case studies we report, etc.

	\subsection{Limitations and Future Work} 
	Naturally, our system is not free from limitations.
	\system relies on language modeling (i.e., TF-IDF scores) to characterize communities; as discussed in Section~\ref{sec:cross}, in case of high overlap in language across two communities, the system can misattribute a given aggression attack.
	\system also requires a certain number of comments to be made on a video before it considers it an aggression attack.
	Therefore, a relatively niche video with a small number of comments might not be considered by our system even though an attack might be taking place.

	\system attributes attacks with an accuracy of 75\%.
	Although our experiments show low false positive rates (at most 0.07), we envision \system being used to flag potential attacks that can be assessed manually or processed further.
	In other words, \system can act as a helpful warning indicator for a potential attack, helping to quickly identify and deter an orchestrated aggression attack.
	
	We believe that \system can also be improved in several ways.
	First, \system relies on TF-IDF to attribute text to source communities;  in the future, more expressive models can be used, e.g., embeddings, to better understand the context in which these words are used.
	Another interesting area of research could be investigating the characteristic patterns of the accounts that engage in coordinated aggression attacks and creating tools that detect and flag accounts that belong to organized campaigns.
	Also, if a video gets traffic from a given community, it does not automatically mean an aggression attack is occurring; thus, future models could incorporate additional factors like the presence of hate speech and the toxicity of comments posted on the video as additional markers of a hate attack. 
	
	Finally, we select \donald, 4chan's~\pol board, and Incel subreddits for our research, partly due to their toxicity and tendency towards targeted hate attacks.
	However, future work should also examine online aggression attacks originating from other communities, perhaps from the other side of the ``political spectrum,'' since the communities we study in this paper are typically associated with the far-right.
	
	\descr{Acknowledgments.} 
	This work was supported by the NSF through grants CNS-1942610, CNS-2114407, and CNS-2127232 and REPHRAIN: the UK's National Research Centre on Privacy, Harm Reduction, and Adversarial Influence Online (UKRI grant EP/V011189/1).

	\small
	\bibliographystyle{apalike}
	%

\begin{thebibliography}{}

\bibitem[Agarwal and Sureka, 2014]{agarwal2014crawler}
Agarwal, S. and Sureka, A. (2014).
\newblock A focused crawler for mining hate and extremism promoting videos on
  youtube.
\newblock In {\em ACM HyperText}.

\bibitem[Ali et~al., 2021]{ali2021understanding}
Ali, S., Saeed, M.~H., Aldreabi, E., Blackburn, J., De~Cristofaro, E.,
  Zannettou, S., and Stringhini, G. (2021).
\newblock Understanding the effect of deplatforming on social networks.
\newblock In {\em ACM WebSci}.

\bibitem[Alshamrani et~al., 2020]{alshamrani2020investigating}
Alshamrani, S., Abuhamad, M., Abusnaina, A., and Mohaisen, D. (2020).
\newblock Investigating online toxicity in users interactions with the
  mainstream media channels on youtube.
\newblock In {\em CIKM (Workshops)}.

\bibitem[Bailey, 2021]{kiwifarms2021}
Bailey, K. (2021).
\newblock {Near, the Programmer Behind the Legendary BSNES Emulator, Has Died}.
\newblock \url{https://www.ign.com/articles/near-bsnes-remembrance}.

\bibitem[Blais and Dupuis‐Déri, 2012]{melissa2012masc}
Blais, M. and Dupuis‐Déri, F. (2012).
\newblock Masculinism and the antifeminist countermovement.
\newblock {\em Social Movement Studies}, 11.

\bibitem[Chandrasekharan et~al., 2019]{chandrasekharan2019crossmod}
Chandrasekharan, E., Gandhi, C., Mustelier, M.~W., and Gilbert, E. (2019).
\newblock Crossmod: A cross-community learning-based system to assist reddit
  moderators.
\newblock {\em Proceedings of the ACM on human-computer interaction}, 3(CSCW).

\bibitem[Chatzakou et~al., 2017]{despoina2017websci}
Chatzakou, D., Kourtellis, N., Blackburn, J., De~Cristofaro, E., Stringhini,
  G., and Vakali, A. (2017).
\newblock Hate is not binary: Studying abusive behavior of \#gamergate on
  twitter.
\newblock In {\em {ACM WebSci}}.

\bibitem[Chelmis and Yao, 2019]{chelmis2019cyber}
Chelmis, C. and Yao, M. (2019).
\newblock Minority report: Cyberbullying prediction on instagram.
\newblock In {\em ACM WebSci}.

\bibitem[Chen et~al., 2019]{chen2019abuse}
Chen, H., McKeever, S., and Delany, S.~J. (2019).
\newblock The use of deep learning distributed representations in the
  identification of abusive text.
\newblock {\em ICWSM}, 13(01):125--133.

\bibitem[Cheng et~al., 2015]{cheng2021anti}
Cheng, J., Danescu-Niculescu-Mizil, C., and Leskovec, J. (2015).
\newblock Antisocial behavior in online discussion communities.
\newblock In {\em ICWSM}.

\bibitem[Collins, 2020]{donaldhatesuspend}
Collins, K. (2020).
\newblock {Reddit bans controversial forum The\_Donald as it revises hate
  speech policies}.
\newblock
  \url{https://www.cnet.com/tech/services-and-software/reddit-bans-controversial-forum-the-donald-as-it-revises-hate-speech-policies/}.

\bibitem[Dadvar et~al., 2014]{dadvar2014expert}
Dadvar, M., Trieschnigg, D., and Jong, F.~d. (2014).
\newblock {Experts and machines against bullies: A hybrid approach to detect
  cyberbullies}.
\newblock In {\em Canadian Conference on Artificial Intelligence}.

\bibitem[Davidson et~al., 2017]{davidson2017AutomatedHateSpeech}
Davidson, T., Warmsley, D., Macy, M., and Weber, I. (2017).
\newblock Automated {{Hate Speech Detection}} and the {{Problem}} of
  {{Offensive Language}}.
\newblock In {\em ICWSM}.

\bibitem[Devlin et~al., 2019]{devlin2019bert}
Devlin, J., Chang, M.-W., Lee, K., and Toutanova, K. (2019).
\newblock {BERT}: Pre-training of deep bidirectional transformers for language
  understanding.
\newblock In {\em NAACL}.

\bibitem[Djuric et~al., 2015]{djuric2015hate}
Djuric, N., Zhou, J., Morris, R., Grbovic, M., Radosavljevic, V., and
  Bhamidipati, N. (2015).
\newblock Hate speech detection with comment embeddings.
\newblock In {\em The Web Conference (WWW)}.

\bibitem[ElSherief et~al., 2018]{mai2018hate}
ElSherief, M., Nilizadeh, S., Nguyen, D., Vigna, G., and Belding, E.~M. (2018).
\newblock Peer to peer hate: Hate speech instigators and their targets.
\newblock {\em arXiv:1804.04649}.

\bibitem[Flores-Saviaga et~al., 2018]{flores2018mobilizing}
Flores-Saviaga, C., Keegan, B.~C., and Savage, S. (2018).
\newblock Mobilizing the trump train: Understanding collective action in a
  political trolling community.
\newblock In {\em ICWSM}.

\bibitem[Giannakopoulos et~al., 2010]{giann2010violence}
Giannakopoulos, T., Pikrakis, A., and Theodoridis, S. (2010).
\newblock {A Multimodal Approach to Violence Detection in Video Sharing Sites}.
\newblock In {\em ICPR}.

\bibitem[Ging, 2019]{debbie2019masc}
Ging, D. (2019).
\newblock Alphas, betas, and incels: Theorizing the masculinities of the
  manosphere.
\newblock {\em Men and Masculinities}, 22(4):638--657.

\bibitem[Google, 2020]{perspectiveAPI}
Google (2020).
\newblock {Perspective API}.
\newblock \url{https://www.perspectiveapi.com}.

\bibitem[Grigg, 2010]{grigg2010cyber}
Grigg, D.~W. (2010).
\newblock {Cyber-Aggression: Definition and Concept of Cyberbullying}.
\newblock {\em Australian Journal of Guidance and Counselling}, 20(2).

\bibitem[Hernandez et~al., 2018]{hernandez2018fraud}
Hernandez, N., Rahman, M., Recabarren, R., and Carbunar, B. (2018).
\newblock Fraud de-anonymization for fun and profit.
\newblock In {\em ACM CCS}.

\bibitem[Hine et~al., 2017]{hine2017kek}
Hine, G.~E., Onaolapo, J., {De Cristofaro}, E., Kourtellis, N., Leontiadis, I.,
  Samaras, R., Stringhini, G., and Blackburn, J. (2017).
\newblock {Kek, Cucks, and God Emperor Trump: A Measurement Study of 4chan's
  Politically Incorrect Forum and Its Effects on the Web}.
\newblock In {\em ICWSM}.

\bibitem[Jaki et~al., 2019]{jaki2019online}
Jaki, S., De~Smedt, T., Gw{\'o}{\'z}d{\'z}, M., Panchal, R., Rossa, A., and
  De~Pauw, G. (2019).
\newblock {Online hatred of women in the Incels.me forum: Linguistic analysis
  and automatic detection}.
\newblock {\em Journal of Language Aggression and Conflict}, 7(2).

\bibitem[Jhaver et~al., 2022]{jhaver2022designing}
Jhaver, S., Chen, Q.~Z., Knauss, D., and Zhang, A.~X. (2022).
\newblock Designing word filter tools for creator-led comment moderation.
\newblock In {\em ACM CHI}.

\bibitem[Kumar et~al., 2018]{kumar2018community}
Kumar, S., Hamilton, W.~L., Leskovec, J., and Jurafsky, D. (2018).
\newblock Community interaction and conflict on the web.
\newblock In {\em The Web Conference (WWW)}.

\bibitem[Kwon and Gruzd, 2017]{kwon2017trump}
Kwon, K. and Gruzd, A. (2017).
\newblock Is offensive commenting contagious online? examining public vs.
  interpersonal swearing in response to donald trump’s youtube campaign
  videos.
\newblock {\em Internet Research}, 27.

\bibitem[Lample and Conneau, 2019]{lample2019cross}
Lample, G. and Conneau, A. (2019).
\newblock Cross-lingual language model pretraining.
\newblock In {\em NeurIPS}.

\bibitem[Liu, 2019]{yang2019bert}
Liu, Y. (2019).
\newblock Fine-tune {BERT} for extractive summarization.
\newblock {\em arXiv:1903.10318}.

\bibitem[Marathe and Shirsat, 2015]{marathe2015approach}
Marathe, S. and Shirsat, P. (2015).
\newblock Approaches for mining youtube videos metadata in cyber bullying
  detection.
\newblock {\em International Journal of Engineering Research and Technology},
  4.

\bibitem[Mariconti et~al., 2019]{mariconti2018you}
Mariconti, E., Suarez-Tangil, G., Blackburn, J., De~Cristofaro, E., Kourtellis,
  N., Leontiadis, I., Serrano, J.~L., and Stringhini, G. (2019).
\newblock {``You Know What to Do'': Proactive Detection of YouTube Videos
  Targeted by Coordinated Hate Attacks}.
\newblock In {\em ACM CSCW}.

\bibitem[Moor et~al., 2010]{moor2010flaming}
Moor, P.~J., Heuvelman, A., and Verleur, R. (2010).
\newblock Flaming on youtube.
\newblock {\em Computers in Human Behavior}, 26(6).

\bibitem[Nilizadeh et~al., 2017]{nilizadeh2017poised}
Nilizadeh, S., Labr{\`e}che, F., Sedighian, A., Zand, A., Fernandez, J.,
  Kruegel, C., Stringhini, G., and Vigna, G. (2017).
\newblock Poised: Spotting twitter spam off the beaten paths.
\newblock In {\em ACM CCS}.

\bibitem[Olteanu et~al., 2018]{alex2018hate}
Olteanu, A., Castillo, C., Boy, J., and Varshney, K.~R. (2018).
\newblock The effect of extremist violence on hateful speech online.
\newblock {\em arXiv:1804.05704}.

\bibitem[Ottoni et~al., 2018]{ottoni2018right}
Ottoni, R., Cunha, E., Magno, G., Bernardina, P., Jr., W.~M., and Almeida, V.
  A.~F. (2018).
\newblock Analyzing right-wing youtube channels: Hate, violence and
  discrimination.
\newblock {\em arXiv:1804.04096}.

\bibitem[Pacheco et~al., 2021]{pacheco2021coord}
Pacheco, D., Hui, P.-M., Torres-Lugo, C., Truong, B., Flammini, A., and
  Menczer, F. (2021).
\newblock Uncovering coordinated networks on social media: Methods and case
  studies.
\newblock {\em ICWSM}, pages 455--466.

\bibitem[Papadamou et~al., 2021]{papa2021over}
Papadamou, K., Zannettou, S., Blackburn, J., De~Cristofaro, E., Stringhini, G.,
  and Sirivianos, M. (2021).
\newblock {``How over is it?'' Understanding the Incel Community on YouTube}.
\newblock {\em Proceedings of the ACM on Human Computer Interaction}, 5(CSCW2).

\bibitem[Papadamou et~al., 2022]{papadamou2022just}
Papadamou, K., Zannettou, S., Blackburn, J., De~Cristofaro, E., Stringhini, G.,
  and Sirivianos, M. (2022).
\newblock ``it is just a flu'': assessing the effect of watch history on
  youtube’s pseudoscientific video recommendations.
\newblock In {\em ICWSM}, pages 723--734.

\bibitem[Papasavva et~al., 2020]{papasavva2020raiders}
Papasavva, A., Zannettou, S., De~Cristofaro, E., Stringhini, G., and Blackburn,
  J. (2020).
\newblock {Raiders of the Lost Kek: 3.5 Years of Augmented 4chan Posts from the
  Politically Incorrect Board}.
\newblock In {\em ICWSM}.

\bibitem[Porter, 2006]{porter_stemmer}
Porter, M. (2006).
\newblock {The Porter Stemming Algorithm}.
\newblock \url{https://tartarus.org/martin/PorterStemmer/}.

\bibitem[Reimers and Gurevych, 2019]{reimer2019sbert}
Reimers, N. and Gurevych, I. (2019).
\newblock Sentence-{BERT}: Sentence embeddings using {S}iamese {BERT}-networks.
\newblock In {\em EMNLP-IJCNLP}.

\bibitem[{Rewire}, 2023]{rewire2023}
{Rewire} (2023).
\newblock {https://rewire.online/}.

\bibitem[Ribeiro et~al., 2021]{Ribeiro2020FromPA}
Ribeiro, M.~H., Blackburn, J., Bradlyn, B., De~Cristofaro, E., Stringhini, G.,
  Long, S., Greenberg, S., and Zannettou, S. (2021).
\newblock The evolution of the manosphere across the web.
\newblock In {\em ICWSM}.

\bibitem[Ribeiro et~al., 2018]{horta2018char}
Ribeiro, M.~H., Calais, P.~H., Santos, Y.~A., Almeida, V.~A., and Meira~Jr, W.
  (2018).
\newblock Characterizing and detecting hateful users on twitter.
\newblock In {\em ICWSM}.

\bibitem[Salminen et~al., 2018]{salminem2019hate}
Salminen, J., Almerekhi, H., Milenkovi{\'c}, M., Jung, S.-g., An, J., Kwak, H.,
  and Jansen, B.~J. (2018).
\newblock {Anatomy of online hate: developing a taxonomy and machine learning
  models for identifying and classifying hate in online news media}.
\newblock In {\em ICWSM}.

\bibitem[Sharma et~al., 2021]{sharma2021coord}
Sharma, K., Zhang, Y., Ferrara, E., and Liu, Y. (2021).
\newblock Identifying coordinated accounts on social media through hidden
  influence and group behaviours.
\newblock In {\em ACM KDD}.

\bibitem[Stringhini et~al., 2010]{stringhini2010detecting}
Stringhini, G., Kruegel, C., and Vigna, G. (2010).
\newblock Detecting spammers on social networks.
\newblock In {\em ACSAC}.

\bibitem[Stringhini et~al., 2015]{stringhini2015evilcohort}
Stringhini, G., Mourlanne, P., Jacob, G., Egele, M., Kruegel, C., and Vigna, G.
  (2015).
\newblock Evilcohort: Detecting communities of malicious accounts on online
  services.
\newblock In {\em USENIX Security Symposium}.

\bibitem[Sureka et~al., 2010]{sureka2010mining}
Sureka, A., Kumaraguru, P., Goyal, A., and Chhabra, S. (2010).
\newblock Mining youtube to discover extremist videos, users and hidden
  communities.
\newblock In {\em Asia Information Retrieval Symposium}.

\bibitem[Tahir et~al., 2019]{tahir2019youtube}
Tahir, R., Ahmed, F., Saeed, H., Ali, S., Zaffar, F., and Wilson, C. (2019).
\newblock Bringing the kid back into youtube kids: Detecting inappropriate
  content on video streaming platforms.
\newblock In {\em ASONAM}.

\bibitem[Tucker-McLaughlin, 2013]{tucker2013youtube}
Tucker-McLaughlin, M. (2013).
\newblock Youtube’s most-viewed videos: Where the girls aren’t.
\newblock {\em Women \& Language}, 36(1).

\bibitem[Weaver et~al., 2012]{weaver2012viol}
Weaver, A.~J., Zelenkauskaite, A., and Samson, L. (2012).
\newblock The (non)violent world of youtube: Content trends in web video.
\newblock {\em Journal of Communication}, 62(6).

\bibitem[Wikipedia, 2022a]{cohen2022}
Wikipedia (2022a).
\newblock {Cohen's kappa}.
\newblock \url{https://en.wikipedia.org/wiki/Cohens_kappa}.

\bibitem[Wikipedia, 2022b]{kstest2022}
Wikipedia (2022b).
\newblock {Kolmogorov-Smirnov test}.
\newblock \url{https://en.wikipedia.org/wiki/Kolmogorov-Smirnov_test}.

\bibitem[Wotanis and McMillan, 2014]{wotanis2014gender}
Wotanis, L. and McMillan, L. (2014).
\newblock Performing gender on youtube.
\newblock {\em Feminist Media Studies}, 14(6).

\bibitem[{YouTube Developers}, 2022]{youtubedatapi}
{YouTube Developers} (2022).
\newblock {YouTube Data API}.
\newblock {https://developers.google.com/youtube/v3/}.

\bibitem[Zach, 2021]{bonferroni2021}
Zach (2021).
\newblock {The Bonferroni Correction Definition \& Example}.
\newblock \url{https://www.statology.org/bonferroni-correction/}.

\bibitem[Zannettou et~al., 2020]{baumgartner2020pushshift}
Zannettou, S., Baumgartner, J., Keegan, B., Squire, M., and Blackburn, J.
  (2020).
\newblock {The Pushshift Reddit Dataset}.
\newblock In {\em ICWSM}.

\bibitem[Zannettou et~al., 2018]{zannettou2018origins}
Zannettou, S., Caulfield, T., Blackburn, J., De~Cristofaro, E., Sirivianos, M.,
  Stringhini, G., and Suarez-Tangil, G. (2018).
\newblock {On the Origins of Memes by Means of Fringe Web Communities}.
\newblock In {\em ACM IMC}.

\bibitem[Zhao et~al., 2009]{zhao2009botgraph}
Zhao, Y., Xie, Y., Yu, F., Ke, Q., Yu, Y., Chen, Y., and Gillum, E. (2009).
\newblock Botgraph: Large scale spamming botnet detection.
\newblock In {\em USENIX Security Symposium}.

\end{thebibliography}

\end{document}